\documentclass[a4paper]{article}

\usepackage{amsmath}
\usepackage{amssymb}
\usepackage{amsthm}
\usepackage{indentfirst}

\makeatletter
\newcommand{\rmnum}[1]{\romannumeral #1}
\newcommand{\Rmnum}[1]{\uppercase\expandafter{\romannumeral #1}}

\newtheorem{thm}{Theorem}
\newtheorem{cor}[thm]{Corollary}
\newtheorem{lem}[thm]{Lemma}
\newtheorem{prop}[thm]{Proposition}

\begin{document}

\title{\textbf{A Remark of the Sanders-Wang's Theorem on Symmetry-integrability}}
\author{Chen Li-Zhou\thanks{E-mail: 031018007@fudan.edu.cn}
    \medskip\\
    \small Institute of Mathematics, Fudan University, Shanghai 200433, China}
\date{}
\maketitle

\noindent{\it\textbf{Abstract}}
  {\small{\it
We extend the integrability analysis for scalar evolution
equations of type
$$u_t=u_m+f(u,u_1,\ldots,u_{m-1})$$
from the case that the right hand side is a $\lambda$-homogeneous
formal power series to the case that it is a nonhomogeneous formal
power series. It is proved that the existence of one nontrivial
symmetry implies the existence of infinitely many, more precisely,
the orders of the infinite integrable hierarchy must be one of the
following cases: $\mathbb{Z}_++1$, $2\mathbb{Z}_++1$,
$6\mathbb{Z}_+\pm1$, or $6\mathbb{Z}_++1$. Moreover, if the
nonlinear part of the equation is a polynomial of order less than
$m-1$, we show that any generalized symmetry is also of polynomial
type.}

  \bigskip
  }

\section{Introduction}
Stimulated by the great progress of the theory of solitons and
integrable systems, the symmetry aspect of PDE systems, which was
initiated by Sophus Lie more than one hundred years ago, has been
intensively studied by many famous researchers in the last three
decades and is still very important. We refer to the book
\cite{Olv93} for history remarks and various applications. An
interesting problem arising in this period is whether a system of
partial differential equations can admit only a finite dimensional
space of generalized symmetries. It is a common knowledge that an
integrable evolution equation is always a member of an infinite
integrable hierarchy whose members are symmetries one for another,
as Fokas \cite{Fok80} stated
\begin{quote}
Another interesting fact regarding the symmetry structure of
evolution equations is that in all known cases the existence of
one generalized symmetry implies the existence of infinitely many.
(However, this has not been proved in general.)
\end{quote}

For $\lambda$-homogeneous, with positive $\lambda$, equations of
the form
$$u_t=u_m+f(u,u_1,\ldots,u_{m-1}),$$
where $f$ is a formal power series with terms that are at least
quadratic, the conjecture has been proved by Sanders and Wang
\cite{SW98}. The exact statement is
\begin{quote}
A nontrivial symmetry of a $\lambda$-homogeneous equation is part
of a hierarchy staring at order 3, 5, or 7 in the odd case, and at
order 2 in the even case.
\end{quote}
Note that $\lambda$-homogeneity with positive $\lambda$ implies
the equation is of polynomial type and is very restricted. The aim
of the present paper is to remove the condition of
$\lambda$-homogeneity and show that the orders of the infinite
integrable hierarchy must be one of the following cases,
Theorem~\ref{one symmetry implies symmetry-integrability}, :
(\rmnum{1}) all positive integers, as the Burgers equation;
(\rmnum{2}) all odd positive integers, as the KdV equation;
(\rmnum{3}) all positive integers congruent to 1 or $-1$ modulo 6,
as the potential Sawada-Kotera equation; (\rmnum{4}) all positive
integers congruent to 1 modulo 6. In the last case, however, no
example has been found as far as we know, cf.\ \cite{vdKST02}.
Furthermore, we prove that if the nonlinear part of the equation
is a polynomial of order less than $m-1$, then any generalized
symmetry is also of polynomial type, Theorem~\ref{differential
polynomial symmetry}.

In contrast to the scalar case, the Fokas' conjecture for systems
of evolution equations has been disproved. An example due to
Bakirov of a fourth order system of two coupled evolution
equations is proved to possess only one nontrivial symmetry of
order six by Beukers, Sanders and Wang \cite{BSW98}. Even for the
refined version of the conjecture \cite{Fok87} that a system of
$m$ evolution equations requires $m$ higher order symmetries in
order to be integrable, a counterexample is given by van~der~Kamp
and Sanders \cite{vdKS02}.

The outline of the paper is as follows. In Section 2, we introduce
some definitions and notations used throughout this paper. In
Section 3, we estimate the orders of homogeneous components of
generalized symmetries by induction on degrees. In particular for
a polynomial evolution equation without submaximal order terms, we
obtain an upper bound for the degree of the symmetry with a
prescribed linear term. Section 4 contains a proof of the main
theorem claimed above. For the reader's convenience, we provide
two appendices which state in our notation some well known results
necessary to understand the text.

The present paper is only a very restricted study of the symmetry
structure of scalar evolution equations. We apologize to whom have
read throughout it and still not found anything, especially
practical examples or physical applications, they are interested
in.

\section{Basic definitions and notations}
Let $\mathbb{R}[u,u_1,u_2,\ldots]$ denote the polynomial ring of
infinitely many variables $u=u_0,\ u_1,\ u_2,\ \ldots$ with real
coefficients (any fixed element of $\mathbb{R}[u,u_1,u_2,\ldots]$
involves only finitely many variables). Elements of
$\mathbb{R}[u,u_1,u_2,\ldots]$ are also called differential
polynomials when $u$ is understood as a function of $x$ and $u_i$
is the $i$th order derivative, $i=0,1,2,\ldots$, with respect to
$x$. And let $\mathbb{R}[[u,u_1,u_2,\ldots]]$ denote the ring of
formal power series in variables $u,u_1,u_2,\ldots$ with real
coefficients (any fixed element of
$\mathbb{R}[[u,u_1,u_2,\ldots]]$ may depend on infinitely many
variables, but its homogeneous components all live in
$\mathbb{R}[u,u_1,u_2,\ldots]$). For $k=1,2,\ldots$,
$\mathcal{M}^k$ will stand for the subset of
$\mathbb{R}[[u,u_1,u_2,\ldots]]$ consists of elements whose
homogeneous components of degree less than $k$ all vanish. For
convenience, if $f\in\mathbb{R}[[u,u_1,u_2,\ldots]]$, write $f^k$
for the $k$th degree homogeneous component of $f$.

If $f\in\mathbb{R}[u,u_1,u_2,\ldots]$ and $f$ is not a constant,
define the order of $f$ is the maximal integer $l$ such that $u_l$
appears in the expression of $f$. The nonzero constants are
regarded as being of order zero. And define the order of the zero
element is $-\infty$. If $f\in\mathbb{R}[[u,u_1,u_2,\ldots]]$, the
order of $f$ is defined to be the maximum value of the orders of
its homogeneous components, and may be $+\infty$. Denote $O(l)$
the subset of $\mathbb{R}[[u,u_1,u_2,\ldots]]$ consists of all
elements of order less than or equal to $l$. Note that $O(l)$ is
well defined for arbitrary real number $l$. In the light of Lemma
\ref{estimates of orders} in section 3 where $d$ is a real number,
which leads to a short proof of Theorem \ref{differential
polynomial symmetry}, we will work freely in the context of real
numbers even though in principle nonnegative integers are
sufficient. By definition, $O(l)=\{0\}$ when $l$ is negative, and
$O(l)$ forms a linear space for any real number $l$.

Since only the autonomous equations are concerned in this paper,
the total derivative operator becomes
$$D=\sum_{i=0}^\infty u_{i+1}\frac{\partial}{\partial u_i},$$
which is well defined on $\mathbb{R}[[u,u_1,u_2,\ldots]]$ because
$D$ preserves the set of homogeneous differential polynomials of a
fixed degree. Let $F,G\in\mathbb{R}[u,u_1,u_2,\ldots]$, define
$$\{F,G\}=\mathbf{v}_{_F}G-\mathbf{v}_{_G}F,$$
where
$$
\mathbf{v}_{_F}=\sum_{i=0}^\infty D^iF\frac{\partial}{\partial
u_i},\quad\mathbf{v}_{_G}=\sum_{i=0}^\infty
D^iG\frac{\partial}{\partial u_i}.
$$
It is well defined because the orders both of $F$ and of $G$ are
finite. In particular, if both $F$ and $G$ are homogeneous
differential polynomials, say, of degree $k$ and $l$ respectively,
then $\{F,G\}$ is a homogeneous differential polynomial of degree
$k+l-1$. Hence we can extend the definition of the bracket onto
$\mathbb{R}[[u,u_1,u_2,\ldots]]$ by defining
$$\{F,G\}=\sum_{s=0}^\infty \sum_{k+l-1=s}\left\{F^k,G^l\right\}.$$
The bracket $\{\ ,\ \}$ is a Lie structure. In fact, the Jacobi
identity can be easily checked using the equality
$[D,\mathbf{v}_{_F}]=0$.

Two evolution equations $u_t=F$ and $u_t=G$, where $F$ and $G$ are
both of finite order, are called ($t$-independent) symmetries of
each other if $\{F,G\}=0$. We will not work out the definition of
generalized symmetries in its most generality. For our purpose,
the formalism derived so far is enough. The reason to adopt the
notation $\{\ ,\ \}$ instead of $[\ ,\ ]$ is that, as pointed out
by A.M.~Vinogradov, see page 10 in \cite{Vin01}, the bracket
between generalized symmetries of scalar equations coincides with
the standard Poisson bracket for first order differential
functions which do not depend on $u$.

\section{Estimates of orders}
The following lemma is the key observation of this section.

\begin{lem}\label{preserve orders}
Let $m,n$ be nonnegative integers, $m\geqslant2$ and
$G\in\mathcal{M}^2$. Then
$$G\in O(n)\iff\{u_m,G\}\in O(m+n-1).$$
\end{lem}

\begin{proof}
The assertion is trivial for $G=0$. Now assume $G$ is nonzero and
the order of $G$ is $n\geqslant0$, it suffices to show that the
order of $\{G,u_m\}$ is exact $m+n-1$. We will prove it using the
formula
$$
\frac{\partial}{\partial u_j}D^m=\sum_{i=0}^{\min\{j,m\}}
\binom{m}{i}D^{m-i}\frac{\partial}{\partial u_{j-i}}.
$$
By definition,
$$
\{G,u_m\}=D^mG-\sum_{i=0}^nu_{m+i}\frac{\partial
G}{\partial u_i}.
$$
It is obvious that $\{G,u_m\}\in O(m+n)$. Since $m\geqslant2$, we
have
$$
\frac{\partial}{\partial u_{m+n}}\{G,u_m\}=\frac{\partial
G}{\partial u_n}-\frac{\partial G}{\partial u_n}=0
$$
and
$$
\frac{\partial}{\partial u_{m+n-1}}\{G,u_m\}=\frac{\partial
G}{\partial u_{n-1}}+mD\frac{\partial G}{\partial u_n}-
\frac{\partial G}{\partial u_{n-1}}=mD\frac{\partial G}{\partial
u_n},
$$
where we have adopted the convention $\frac{\partial}{\partial
u_{-1}}G=0$. $G$ is of $n$th order and $G\in\mathcal{M}^2$ imply
$\frac{\partial G}{\partial u_n}$ is not a constant, and hence
$D\frac{\partial G}{\partial u_n}$ is nonzero. Therefore, the
order of $\{G,u_m\}$ is $m+n-1$.
\end{proof}

\begin{lem}\label{estimates of orders}
Suppose $F,G\in\mathcal{M}^1$, $F^1=u_m$, $G^1=u_n$, where
$m,n\geqslant2$. And suppose $\{F,G\}\in\mathcal{M}^{s_0+1}$, for
some $s_0\geqslant2$. If there is a real number $d$ such that
$F^k\in O(m-1-(k-1)d),\,\forall\,k,2\leqslant k\leqslant s_0$,
then
$$G^l\in O(n-1-(l-1)d),\,\forall\,l,2\leqslant l\leqslant s_0.$$
\end{lem}

\begin{proof}
Since $\mathcal{M}^{s_0+1}\subset\mathcal{M}^{s_0}$, by induction
on $s_0$, we may assume
$$G^l\in O(n-1-(l-1)d),\ \forall\ l, 2\leqslant l\leqslant s_0-1.$$
Observe that $\{O(m),O(n)\}\subset O(m+n)$ holds for arbitrary
real numbers $m,n$ as well as for nonnegative integers and that
Lemma~\ref{preserve orders} holds for arbitrary real number $n$ as
well as for nonnegative integer $n$. From
$\{F,G\}^{s_0}=\sum_{k+l-1=s_0}\left\{F^k,G^l\right\}=0$, we have
\begin{equation*}
\begin{split}
\left\{G^{s_0},u_m\right\} &
=\sum_{k=2}^{s_0-1}\left\{F^k,G^{s_0-k+1}\right\}+\left\{F^{s_0},u_n\right\} \\
& \in\sum_{k=2}^{s_0-1}\{O(m-1-(k-1)d),O(n-1-(s_0-k)d)\} \\[6pt] & \quad
{}+\{O(m-1-(s_0-1)d),u_n\} \\[6pt] & \subset O(m+n-2-(s_0-1)d).
\end{split}
\end{equation*}
By Lemma~\ref{preserve orders}, $G^{s_0}\in O(n-1-(s_0-1)d)$.
\end{proof}

Lemma~\ref{estimates of orders} serves two purposes. Consider the
equation $u_t=F=u_m+f$ where $f\in\mathcal{M}^2\cap O(m-1)$.
Suppose we have obtained a solution $G=u_n+g$ of the symmetry
equation $\{F,G\}=0$, where $g$ lives in $\mathcal{M}^2$, then
applying Lemma~\ref{estimates of orders} for the case $d=0$, we
get $g\in O(n-1)$. This is the last step of the main theorem in
the next section. And the case $d>0$ of Lemma~\ref{estimates of
orders} leads to the following

\begin{thm}\label{differential polynomial symmetry}
Suppose $F=u_m+f$, $G=u_n+g$, where $m,n\geqslant2$,
$f,g\in\mathcal{M}^2$. If $\{F,G\}=0$ and if $f$ is a differential
polynomial of order less than $m-1$, then $g$ is a differential
polynomial of order less than $n-1$.
\end{thm}

\begin{proof}
Since $f$ is a differential polynomial of order less than $m-1$,
it is easy to see that $f^k\in
O(m-1-(k-1)d),\,\forall\,k\geqslant2$, for sufficient small $d>0$.
By Lemma \ref{estimates of orders}, $g^l\in
O(n-1-(l-1)d),\,\forall\,l\geqslant2$. Hence the order of $g$ is
less than $n-1$. When $n-1-(l-1)d<0$, $g^l=0$. Therefore $g$ is
also a differential polynomial and its degree is not bigger than
$\frac{n-1}{d}+1$.
\end{proof}

The upper bound for the degree in the preceding proof may be not
accurate in particular examples. Nevertheless, applying directly
the method of estimating orders we have derived may give sharp
upper bounds for the degrees of generalized symmetries before
their explicit expressions are obtained. For example, suppose
$G=u_{2k+1}+g$, where $g\in\mathcal{M}^2$, is a generalized
symmetry of the KdV equation $u_t=u_3+uu_1$, then estimating
inductively the orders of the homogeneous components of $g$ yields
that $g^l\in O(2(k-l+1)+1)$, hence the degree of $g$ is not bigger
than $k+1$. More generally, for any equation of the form
$u_t=F=u_{2k_0+1}+f$ where $f\in\mathcal{M}^2$ admits the estimate
$f^l\in O(2(k_0-l+1)+1)$, e.g. the potential Sawada-Kotera
equation $u_t=u_5+5u_1u_3+{5\over3}u_1^3$, the same conclusion
follows.

\section{Symmetry-integrability}
We now proceed to prove the theorem claimed before:

\textit{For any scalar evolution equation of the form
$$u_t=u_m+f(u,u_1,\ldots,u_{m-1}),$$
where $f$ is a formal power series with terms that are at least
quadratic, the existence of one nontrivial symmetry implies the
existence of infinitely many. Moreover, the orders of the infinite
integrable hierarchy must be one of the following cases:
(\rmnum{1}) all positive integers; (\rmnum{2}) all odd positive
integers; (\rmnum{3}) all positive integers congruent to 1 or $-1$
modulo 6; (\rmnum{4}) all positive integers congruent to 1 modulo
6.}

Let us begin with a consequence of the Beukers' theorem, see
Appendix 1.

\begin{cor}\label{pull back}
Let $m,n\geqslant2$, $k_0=
\begin{cases}
3, & 2\mid mn\\4, & 2\nmid mn
\end{cases}$,
and $k\geqslant k_0$. And let $F,G$ be two homogeneous
differential polynomials of $k$th degrees satisfying
$\{F,u_n\}=\{G,u_m\}$. Then there exists a unique "pull back" $H$,
also a homogeneous differential polynomial of $k$th degree, s.t.
$F=\{H,u_m\}$ and $G=\{H,u_n\}$.
\end{cor}

\begin{proof}
By the Beukers' theorem (see (\ref{factorization2}) and
(\ref{factorization3})), $P_k^{(m)}$ and $P_k^{(n)}$ are relative
prime. Thanks to the Gel'fand-Diki\u{\i} transformation, we have
$\widetilde{F}P_k^{(n)}=\widetilde{G}P_k^{(m)}$. Hence
$P_k^{(m)}\mid\widetilde{F}$ and $P_k^{(n)}\mid\widetilde{G}$. Set
$\widetilde{H}=\frac{\widetilde{F}}{P_k^{(m)}}=
\frac{\widetilde{G}}{P_k^{(n)}}$, then the preimage $H$ of
$\widetilde{H}$ under the Gel'fand-Diki\u{\i} transformation is
the needed. The uniqueness of $H$ is obvious.
\end{proof}

To avoid endlessly repeating the hypothesis, let us denote
$$
W_l=\left\{u_l+f\left|f\in\mathcal{M}^2\cap O(l-1)\right.\right\}
$$
for arbitrary $l\geqslant2$, and
$$W=\bigcup_{l=2}^\infty W_l.$$
From now on in this section, we always assume $F\in W_m,\ G\in
W_n$, satisfying $\{F,G\}=0$, where $m,n\geqslant2$ and $m\neq n$.
It just means that the equation $u_t=F$ has a nontrivial symmetry
$G$.

\begin{prop}\label{transitive}
Suppose $E\in\mathcal{M}$ and $k\geqslant2$.

$(\rmnum{1})$ If $\{E,F\},\{E,G\}\in\mathcal{M}^k$, then
$\left\{\{E,F\}^k,u_n\right\}=\left\{\{E,G\}^k,u_m\right\}$;

$(\rmnum{2})$ If $\{E,F\}\in\mathcal{M}^{k+1}$ and
$\{E,G\}\in\mathcal{M}^k$, then $\{E,G\}\in\mathcal{M}^{k+1}$;

$(\rmnum{3})$ If $\{E,F\}=0$, then $\{E,G\}=0$.
\end{prop}

\begin{proof}
$(\rmnum{1})$ Since $\{F,G\}=0$, by the Jacobi identity, we have
$$\left\{\{E,F\},G\right\}=\left\{\{E,G\},F\right\}.$$
Taking the $k$th degree homogeneous components of the two sides of
the above equality, we get
$$\left\{\{E,F\}^k,u_n\right\}=\left\{\{E,G\}^k,u_m\right\}.$$

$(\rmnum{2})$ By condition, part $(\rmnum{1})$ holds and
$\{E,F\}^k=0$. Thus $\left\{\{E,G\}^k,u_m\right\}=0$. It, see
Appendix 1, implies $\{E,G\}^k=0$.

$(\rmnum{3})$ It is easy to see that $\{E,G\}\in\mathcal{M}^2$.
The conclusion follows from part $(\rmnum{2})$ by induction since
$\bigcap_{l=2}^\infty\mathcal{M}^l=0$.
\end{proof}

Consider the linear space of nontrivial symmetries of the equation
$u_t=F$
$$\mathcal{F}=\,\mathrm{span}\,\{E\in W|\{E,F\}=0\}$$
and the subspaces of $l$th order symmetries with a single linear
term, $l=2,3,\ldots$,
$$\mathcal{F}_l=\,\mathrm{span}\,W_l\cap\mathcal{F}.$$

We know $\,\dim\,\mathcal{F}_l=0 \mbox{ or }1$, and
$\mathcal{F}=\bigoplus_{l=2}^\infty\mathcal{F}_l$. By Proposition
\ref{transitive} $(\rmnum{3})$, $\mathcal{F}$ is a commutative Lie
subalgebra of $\left(\mathbb{R}[[u,u_1,u_2,\ldots]],\ \{\ ,\
\}\right)$. And $\,\dim\,\mathcal{F}\geqslant2$, since $F$ belongs
to $\mathcal{F}$ and we have assumed the existence of $G$. Now the
main theorem can be reformulated as follows.

\begin{thm}\label{one symmetry implies symmetry-integrability}
The space $\mathcal{F}$ is infinite dimensional. More precisely,
$$
\left\{l\left|\,\dim\,\mathcal{F}_l=1\right.\right\}=
\mathbb{Z}_{>0}+1,\mbox{ or }2\mathbb{Z}_{>0}+1,\mbox{ or
}6\mathbb{Z}_{>0}\pm1,\mbox{ or }6\mathbb{Z}_{>0}+1.
$$
\end{thm}

Here is our key observation. Without losing generality, we may
assume
$$
\left\{l\left|\,\dim\,\mathcal{F}_l=1\right.\right\}\subset
\left\{l\left|\,t_2^{(m)}\mid t_2^{(l)}\right.\right\},
$$
see (\ref{factorization1}). In fact, there exists $F'\in
W_{m'}\cap\mathcal{F}$, such that $t_2^{(m')}\mid t_2^{(l)}$, for
any $l$ satisfying $\,\dim\,\mathcal{F}_l=1$. By Proposition
\ref{transitive} $(\rmnum{3})$, we can replace $F$ by $F'$ without
changing $\mathcal{F}$. And by (\ref{factorization1}),
$$
\left\{l\left|\,t_2^{(m)}\mid t_2^{(l)}\right.\right\}=
\begin{cases}
\mathbb{Z}_{>0}+1, & m=0\bmod 2;\\
2\mathbb{Z}_{>0}+1, & m=3\bmod 6;\\
6\mathbb{Z}_{>0}\pm1, & m=5\bmod 6;\\
6\mathbb{Z}_{>0}+1, & m=1\bmod 6.
\end{cases}
$$
Thus it remains to show
$$
\left\{l\left|\,\dim\,\mathcal{F}_l=1\right.\right\}\supset
\left\{l\left|\,t_2^{(m)}\mid t_2^{(l)}\right.\right\}.
$$

We have reduced Theorem \ref{one symmetry implies
symmetry-integrability} to the following

\begin{thm}\label{reduced part}
Let $l\geqslant2$. If $t_2^{(m)}\mid t_2^{(l)}$, then
$\,\dim\,\mathcal{F}_l=1$.
\end{thm}

\begin{proof}
We shall show that there exists $E\in W_l$ for $l\ne m$, s.t.
$\{E,F\}=0$. First, let $E^1=u_l$.

Taking the second degree homogeneous component of the equality
$\{F,G\}=0$, we get
$\left\{F^2,u_n\right\}=\left\{G^2,u_m\right\}$, equivalently,
$\widetilde{F^2}t_2^{(n)}p_2^{(n)}=
\widetilde{G^2}t_2^{(m)}p_2^{(m)}$. Hence $p_2^{(m)},p_2^{(n)}$
divide $\widetilde{F^2},\widetilde{G^2}$ respectively. Set
\begin{equation}\label{the quadratic term}
\widetilde{E^2}=\frac{\widetilde{F^2}}{p_2^{(m)}}
\frac{t_2^{(l)}}{t_2^{(m)}}p_2^{(l)},
\end{equation}
then $\left\{E^1+E^2,F\right\}^2=0$, i.e.
$\left\{E^1+E^2,F\right\}\in\mathcal{M}^3$.

By Proposition \ref{transitive} $(\rmnum{2})$,
$\left\{E^1+E^2,G\right\}\in\mathcal{M}^3$. Then by Proposition
\ref{transitive} $(\rmnum{1})$,
$\left\{\left\{E^1+E^2,F\right\}^3,u_n\right\}=
\left\{\left\{E^1+E^2,G\right\}^3,u_m\right\}$. It implies that
$p_3^{(m)},p_3^{(n)}$ divide $\widetilde{\left\{E^1,F^3\right\}}+
\widetilde{\left\{E^2,F^2\right\}},\
\widetilde{\left\{E^1,G^3\right\}}+
\widetilde{\left\{E^2,G^2\right\}}$ respectively.

Taking the third degree homogeneous component of the equality
$\{F,G\}=0$, we get
$\left\{F^2,G^2\right\}+\left\{F^3,u_n\right\}=
\left\{G^3,u_m\right\}$, equivalently,
$\widetilde{\left\{F^2,G^2\right\}}+F^3P_3^{(n)}=G^3P_3^{(m)}$.
When $2\nmid mn$, we obtain $(x_1+x_2)(x_2+x_3)(x_3+x_1)\mid
\widetilde{\left\{F^2,G^2\right\}}$, see (\ref{factorization2}).

Now we need another lemma which is the same as Proposition 5.3 in
\cite{SW98}. For the reader's convenience, we provide a proof in
our notation (without referring to $\lambda$-homogeneity) in
Appendix 2.

\begin{lem}\label{quadratic divisibility}
If $2\nmid lm$, then
$$
(x_1+x_2)(x_2+x_3)(x_3+x_1)\mid \widetilde{\left\{E^2,F^2\right\}}
\iff x_1+x_2\mid\widetilde{F^2}\mbox{ or }
x_1x_2\mid\widetilde{F^2}.
$$
\end{lem}

When $2\nmid mn$, using Lemma \ref{quadratic divisibility}, we
obtain $x_1+x_2\mid\widetilde{F^2}$ or $x_1x_2\mid\widetilde{F^2}$
from $t_3^{(m)}\mid\widetilde{\left\{F^2,G^2\right\}}$. Since
$t_2^{(m)}\mid t_2^{(l)}$, $l$ is also odd. Using Lemma
\ref{quadratic divisibility} again, we obtain
$t_3^{(m)}\mid\widetilde{\left\{E^2,F^2\right\}}$. Consequently,
$P_3^{(m)}=t_3^{(m)}p_3^{(m)}$ divides
$\widetilde{\left\{E^1,F^3\right\}}+
\widetilde{\left\{E^2,F^2\right\}}$. Set $\widetilde{E^3}$ be the
quotient of them, then
$\left\{E^1+E^2+E^3,F\right\}\in\mathcal{M}^4$.

In sum, let $k_0=
\begin{cases}
3, & 2\mid mn\\4, & 2\nmid mn
\end{cases}$,
we have obtained
$$
\overline{E}=
\begin{cases}
E^1+E^2, & 2\mid mn\\E^1+E^2+E^3, & 2\nmid mn
\end{cases},
$$
satisfying $\left\{\overline{E},F\right\}\in\mathcal{M}^{k_0}$.

By Proposition \ref{transitive} $(\rmnum{2})$,
$\left\{\overline{E},G\right\}\in\mathcal{M}^{k_0}$. Then by
Proposition \ref{transitive} $(\rmnum{1})$, we see that
$\left\{\left\{\overline{E},F\right\}^{k_0},u_n\right\}=
\left\{\left\{\overline{E},G\right\}^{k_0},u_m\right\}$. Now
applying Corollary \ref{pull back}, there exists a homogeneous
differential polynomial $E^{k_0}$ of degree $k_0$, such that
$\left\{\overline{E},F\right\}^{k_0}=\left\{-E^{k_0},u_m\right\}$
and
$\left\{\overline{E},G\right\}^{k_0}=\left\{-E^{k_0},u_n\right\}$.
Thus $\left\{\overline{E}+E^{k_0},F\right\},
\left\{\overline{E}+E^{k_0},G\right\}$ belong to
$\mathcal{M}^{k_0+1}$.

By induction, we can obtain a formal power series solution $E$ of
the symmetry equation $\{E,F\}=0$ satisfying $E^0=0$ and
$E^1=u_l$. Finally, by the arguments after Lemma \ref{estimates of
orders}, $E\in W_l$.
\end{proof}

\noindent{\it Remark.} In \cite{SW98}, Sanders and Wang have
formulated Proposition \ref{transitive} $(\rmnum{1})$,
$(\rmnum{2})$ and the induction part of the proof of Theorem
\ref{reduced part} in terms of Lie algebraic modules. As we have
seen, however, they are all rather simple and the abstract setting
is not necessary in our context. It is worse that the abstract
setting has concealed Corollary \ref{pull back} and Proposition
\ref{transitive} $(\rmnum{3})$, although they seem to be also very
simple.

\section*{Acknowledgements}
The author is sincerely grateful to Prof. C.H.~Gu and Prof.
H.S.~Hu for their constant support and encouragement. And the
author owe especial thanks to Prof. E.G.~Fan and Prof. Z.X.~Zhou
for helpful discussions.

\section*{Appendix 1: the symbolic method}
The purpose of this appendix is to introduce some basic results of
the symbolic method, which is first introduced by Gel'fand and
Diki\u{\i} in \cite{GD75} and play a key role in Section 4, see
\cite{SW98,TQ81,Beu97} for proofs.

Let $k$ be a natural number. Write $U^k$ for the set of $k$th
degree homogeneous differential polynomials in
$\mathbb{R}[[u,u_1,u_2,\ldots]]$. Denote, as usual,
$\mathbb{R}[x_1,\ldots,x_k]$ for the polynomial ring of variables
$x_1,\ldots,x_k$ with real coefficients and $\Lambda_k$ the set of
symmetric polynomials in $\mathbb{R}[x_1,\ldots,x_k]$. The well
known symmetrizing operator, denoted by $\langle\ \rangle$, from
$\mathbb{R}[x_1,\ldots,x_k]$ to $\Lambda_k$ is defined by
$$
f(x_1,\ldots,x_k)\mapsto\langle f \rangle= \frac{1}{k!}
\sum_{\sigma\in
S_k}f\left(x_{\sigma(1)},\ldots,x_{\sigma(k)}\right),
$$
where $S_k$ denotes the $k$th symmetry group.

The Gel'fand-Diki$\breve{\i}$ transformation is a linear
isomorphism between $U^k$ and $\Lambda_k$. Its action on the
monomials in $U^k$ is as follows
$$
u^\alpha=u^{\alpha_0}u_1^{\alpha_1}\cdots u_m^{\alpha_m}\mapsto
\widetilde{u^\alpha}=\left\langle x_1^0\cdots x_{\alpha_0}^0
x_{\alpha_0+1}^1\cdots x_{\alpha_0+\alpha_1}^1\cdots
x_{k-\alpha_m+1}^m\cdots x_k^m\right\rangle,
$$
where $\sum_{i=0}^m\alpha_i=k$.

For arbitrary $F\in U^k$, $G\in U^l$, we have
\begin{eqnarray*}
& (\rmnum{1}) & \widetilde{DF}=\widetilde{F}\sum_{i=1}^k x_i;\\
& (\rmnum{2}) & \widetilde{\frac{\partial F}{\partial u_m}}=
\left.\frac{k}{m!}\frac{\partial^m\widetilde{F}}{\partial x_k^m}
\right|_{x_k=0}; \\
& (\rmnum{3}) & \widetilde{\mathbf{v}_{_F}G}= l\left\langle
\widetilde{F}\left(x_1,\ldots,x_k\right)\widetilde{G}
\left(\sum_{i=1}^k x_i,\,
x_{k+1},\ldots,x_{k+l-1}\right)\right\rangle, \\
& &
\begin{split}
\widetilde{\{F,G\}}= & l\left\langle
\widetilde{F}\left(x_1,\ldots,x_k\right)\widetilde{G}
\left(\sum_{i=1}^k x_i,\,
x_{k+1},\ldots,x_{k+l-1}\right)\right\rangle
\\& {}- k\left\langle
\widetilde{G}\left(x_1,\ldots,x_l\right)\widetilde{F}
\left(\sum_{j=1}^l x_j,\,
x_{l+1},\ldots,x_{k+l-1}\right)\right\rangle;
\end{split}
\\& (\rmnum{4}) & \widetilde{\{F,u_m\}}= \widetilde{F}P_k^{(m)},
\mbox{ where } P_k^{(m)}=\left(\sum_{i=1}^k x_i\right)^m-
\sum_{i=1}^k x_i^m.
\end{eqnarray*}

When $k,m\geqslant2$, $P_k^{(m)}$ is nonconstant. It immediately
follows from (\rmnum{4}) that if $F\in\mathcal{M}^2$ and
$\{F,u_m\}=0$, $m\geqslant2$, then $F=0$.

\newtheorem*{thm*}{Theorem}
\begin{thm*}[Beukers]
The symmetric polynomials $P_k^{(m)}$'s have factorizations
$P_k^{(m)}=t_k^{(m)}p_k^{(m)}$, such that (the greatest common
divisor)
$$
\gcd\left(t_k^{(m)},p_k^{(n)}\right)=\gcd\left(p_k^{(m)},p_k^{(n)}\right)=1,\
\forall\ k,m,n\geqslant2,
$$
where $t_k^{(m)},p_k^{(m)}\in\Lambda_k$ and $t_k^{(m)}$'s are as
follows.

$\bullet\ k=2:$
\begin{equation}\label{factorization1}
t_2^{(m)}=
\begin{cases}
x_1x_2, & m=0\bmod 2;\\
x_1x_2(x_1+x_2), & m=3\bmod 6;\\
x_1x_2(x_1+x_2)\left(x_1^2+x_1x_2+x_2^2\right), & m=5\bmod 6;\\
x_1x_2(x_1+x_2)\left(x_1^2+x_1x_2+x_2^2\right)^2, & m=1\bmod 6.
\end{cases}
\end{equation}

$\bullet\ k=3:$
\begin{equation}\label{factorization2}
t_3^{(m)}=
\begin{cases}
1, & m=0\bmod 2;\\
(x_1+x_2)(x_2+x_3)\left(x_3+x_1\right), & m=1\bmod 2.
\end{cases}
\end{equation}

$\bullet\ k\geqslant4:$
\begin{equation}\label{factorization3}
t_k^{(m)}=1.
\end{equation}
\end{thm*}

\section*{Appendix 2}
\noindent{\it Proof of Lemma \ref{quadratic divisibility}.} First
of all, note that $x_1+x_2\mid\widetilde{E^2}$ is equivalent to
$x_1+x_2\mid\widetilde{F^2}$ and that $x_1x_2\mid\widetilde{E^2}$
is equivalent to $x_1x_2\mid\widetilde{F^2}$, following from
(\ref{the quadratic term}) in Section 4 and the Beukers' theorem
in Appendix 1.

According to the second formula of (\rmnum{3}) in Appendix 1, we
have
\begin{equation*}
\begin{split}
\widetilde{\left\{E^2,F^2\right\}}= & 2\left\langle
\widetilde{E^2}(x_1,x_2)\widetilde{F^2}(x_1+x_2,x_3) -
\widetilde{F^2}(x_1,x_2)\widetilde{E^2}(x_1+x_2,x_3) \right\rangle\\
= & \frac{2}{3}\Big(
\widetilde{E^2}(x_1,x_2)\widetilde{F^2}(x_1+x_2,x_3) -
\widetilde{F^2}(x_1,x_2)\widetilde{E^2}(x_1+x_2,x_3) \\
& {}+ \widetilde{E^2}(x_2,x_3)\widetilde{F^2}(x_2+x_3,x_1) -
\widetilde{F^2}(x_2,x_3)\widetilde{E^2}(x_2+x_3,x_1) \\
& {}+ \widetilde{E^2}(x_3,x_1)\widetilde{F^2}(x_3+x_1,x_2) -
\widetilde{F^2}(x_3,x_1)\widetilde{E^2}(x_3+x_1,x_2) \Big).
\end{split}
\end{equation*}
Since $\widetilde{\left\{E^2,F^2\right\}}$ is a symmetric
polynomial,
\begin{equation*}
\begin{split}
(x_1+x_2)(x_2+x_3)(x_3+x_1)\mid \widetilde{\left\{E^2,F^2\right\}}
& \iff x_2+x_3\mid\widetilde{\left\{E^2,F^2\right\}}\\
& \iff\widetilde{\left\{E^2,F^2\right\}}(x_1,x_2,-x_2)=0.
\end{split}
\end{equation*}
Observe that
$$
P_2^{(m)}(x_1,x_2)=\left(x_1+x_2\right)^m-x_1^m-x_2^m=-P_2^{(m)}(x_1+x_2,-x_2).
$$
Thus it follows from
$\widetilde{E^2}P_2^{(m)}=\widetilde{F^2}P_2^{(l)}$ that
$\widetilde{E^2}(x_1+x_2,-x_2)P_2^{(m)}(x_1,x_2)=
\widetilde{F^2}(x_1+x_2,-x_2)P_2^{(l)}(x_1,x_2)$. Consequently,
$$
\widetilde{E^2}(x_1,x_2)\widetilde{F^2}(x_1+x_2,-x_2)=
\widetilde{F^2}(x_1,x_2)\widetilde{E^2}(x_1+x_2,-x_2).
$$
Changing the variable $x_2$ to $-x_2$, we get
$$
\widetilde{E^2}(-x_2,x_1)\widetilde{F^2}(-x_2+x_1,x_2)=
\widetilde{F^2}(-x_2,x_1)\widetilde{E^2}(-x_2+x_1,x_2).
$$
Hence
$$
\widetilde{\left\{E^2,F^2\right\}}(x_1,x_2,-x_2)=
\frac{2}{3}\left( \widetilde{E^2}(x_2,-x_2)\widetilde{F^2}(0,x_1)
- \widetilde{F^2}(x_2,-x_2)\widetilde{E^2}(0,x_1)\right).
$$
In addition, since $l,m$ are odd integers, according to
(\ref{factorization1}) and the equality
$\widetilde{E^2}P_2^{(m)}=\widetilde{F^2}P_2^{(l)}$, we have
$$
\widetilde{E^2}(x_2,x_3)\frac{P_2^{(m)}(x_2,x_3)}{x_2+x_3}=
\widetilde{F^2}(x_2,x_3)\frac{P_2^{(l)}(x_2,x_3)}{x_2+x_3}
$$
and
$$
\widetilde{E^2}(x_2+x_3,x_1)\frac{P_2^{(m)}(x_2+x_3,x_1)}{x_2+x_3}=
\widetilde{F^2}(x_2+x_3,x_1)\frac{P_2^{(l)}(x_2+x_3,x_1)}{x_2+x_3}.
$$
Multiplying  the above two equations by cross and setting
$x_3=-x_2$, we may obtain
\begin{equation*}
\begin{split}
x_2+x_3\mid\widetilde{\left\{E^2,F^2\right\}} \iff &
\widetilde{E^2}(x_2,-x_2)\widetilde{F^2}(0,x_1)=
\widetilde{F^2}(x_2,-x_2)\widetilde{E^2}(0,x_1)\\ \iff &
\widetilde{E^2}(x_2,-x_2)\widetilde{F^2}(0,x_1)=
\widetilde{F^2}(x_2,-x_2)\widetilde{E^2}(0,x_1)=0\\ & \mbox{or
}(x_2+x_3)^3\mid Q(x_1,x_2,x_3),
\end{split}
\end{equation*}
where
$$
Q(x_1,x_2,x_3)=P_2^{(m)}(x_2,x_3)P_2^{(l)}(x_2+x_3,x_1)-
P_2^{(l)}(x_2,x_3)P_2^{(m)}(x_2+x_3,x_1).
$$
But $(x_2+x_3)^3\nmid Q(x_1,x_2,x_3)$, because
\begin{equation*}
\begin{split}
\left.\frac{\partial^2Q}{\partial x_3^2}\right|_{x_3=-x_2}=&
2\bigg( \frac{\partial}{\partial x_3}P_2^{(m)}(x_2,x_3)
\frac{\partial}{\partial x_3}P_2^{(l)}(x_2+x_3,x_1)\\ & {}-
\frac{\partial}{\partial x_3}P_2^{(l)}(x_2,x_3)
\frac{\partial}{\partial x_3}P_2^{(m)}(x_2+x_3,x_1)
\bigg)_{x_3=-x_2}\\ =& 2lm\left(-x_2^{m-1}x_1^{l-1}-
\left(-x_2^{l-1}\right)x_1^{m-1}\right)\ne0.\ (l\ne m)
\end{split}
\end{equation*}
Finally,
\begin{equation*}
\begin{split}
\widetilde{E^2}(x_2,-x_2)\widetilde{F^2}(0,x_1)=0\iff &
\widetilde{E^2}(x_2,-x_2)=0\mbox{ or }\widetilde{F^2}(0,x_1)=0 \\
\iff & x_1+x_2\mid\widetilde{E^2}\mbox{ or }x_2\mid\widetilde{F^2}
\\ \iff & x_1+x_2\mid\widetilde{F^2}\mbox{ or } x_1x_2\mid\widetilde{F^2}.
\end{split}
\end{equation*}
In the same manner,
$$
\widetilde{F^2}(x_2,-x_2)\widetilde{E^2}(0,x_1)=0\iff
x_1+x_2\mid\widetilde{F^2}\mbox{ or } x_1x_2\mid\widetilde{F^2}.
$$

The conclusion follows. \hfill$\Box$

\end{document}